\documentstyle[12pt,psfig]{article}

\newcommand{\vxp}{{\bf x}_\perp}

\newcommand{\beq}{\begin{equation}}
\newcommand{\eeq}[1]{\label{#1} \end{equation}}

\newcommand{\lton}{\mathrel{\lower.9ex\hbox{$\stackrel{\displaystyle 
<}{\sim}$}}}
\newcommand{\ee}{\end{equation}} \newcommand{\ben}{\begin{enumerate}}
\newcommand{\een}{\end{enumerate}} \newcommand{\bit}{\begin{itemize}}
\newcommand{\eit}{\end{itemize}} \newcommand{\bc}{\begin{center}}
\newcommand{\ec}{\end{center}} \newcommand{\bea}{\begin{eqnarray}}
\newcommand{\eea}{\end{eqnarray}}
\newcommand{\beqar}{\begin{eqnarray}}
\newcommand{\eeqar}[1]{\label{#1} \end{eqnarray}}
\newcommand{\vx}{{\bf x}}
\newcommand{\vp}{{\bf p}}

\newcommand{\vpp}{{\bf p}_\perp}

\newcommand{\sh}{{\rm sh}}
\newcommand{\ch}{{\rm ch}}

\newcommand{\de}{\sigma}
\newcommand{\ep}{{\rm e}_\perp}
\newcommand{\dde}{\dot{\sigma}}

\begin{document} 
\begin{flushright}
DOE/ER/40561-334-INT96-21-07\\
CU-TH-857\\[4ex]
\end{flushright}

\begin{center}
{\Large\bf
Transverse Energy Evolution as a Test of Parton Cascade Models
}\footnote[0]{ PREPARED FOR THE U.S. DEPT.  OF ENERGY UNDER GRANTS
    DE-FG06-90ER40561, DE-FG02-93ER40764,
DE-FG-02-92-ER40699, and DE-AC02-76CH00016.
 This report was prepared as an account of work
    sponsored by the United States Government. Neither the United States nor
    any agency thereof, nor any of their employees, makes any warranty, express
    or implied, or assumes any legal liability or responsibility for the
    accuracy, completeness, or usefulness of any information, apparatus,
    product, or process disclosed, or represents that its use would not
    infringe privately owned rights.  Reference herein to any specific
    commercial product, process, or service by trade name, mark, manufacturer,
    or otherwise, does not necessarily constitute or imply its endorsement,
    recommendation, or favoring by the United States Government or any agency
    thereof. The views and opinions of authors expressed herein do not
    necessarily state or reflect those of the United States Government or any
    agency thereof.} \\[3ex]
{ Miklos Gyulassy$^{1,2}$, Yang Pang$^{1,3}$, and Bin Zhang$^{1,2}$}\\[2ex]
{\small 
1.Department of Physics, Columbia University, New York, NY  10027  USA\\
2.INT, University of Washington, Box 351550
Seattle, WA 98195, USA\\
3. Brookhaven National Laboratory, 
Upton, New York, 11973
}\\[2ex]
\date{\today}
\end{center}
\begin{abstract}
We propose a  test of Monte Carlo Parton Cascade models
based on  analytic solutions of covariant kinetic theory 
for longitudinally boost and transverse translation invariant 
boundary conditions. We compute the evolution of the transverse energy per unit
rapidity for typical mini-jet initial conditions expected in
ultra-relativistic nuclear collisions. The kinetic theory
solutions under these conditions test the models
severely because they deviate strongly from  free-streaming
and also from ideal Euler and dissipative Navier-Stokes hydrodynamical
approximations. We show that the newly formulated ZPC model passes this test.
In addition, we show that the initial mini-jet  density 
would need to be approximately
four times higher than estimated  with the HIJING generator in central Au+Au
collisions at c.m. energies 200 AGeV in order that parton cascade dynamics
can be approximated by Navier-Stokes hydrodynamics.
\end{abstract} 

\section{Introduction}

Monte Carlo parton cascade models
\cite{gcp,zpc,Wangmg,vni,uqmd,venus,art,lucea,rqmd} are currently being
developed to calculate multiparticle production and 
evolution in ultra-relativistic nuclear
collisions. These are being written in a new Open Standard Codes and Routines
(OSCAR) format\cite{oscar} to enable rigorous testing of essential components
of the algorithms and insure reproducibility of the numerical results.  Given
the present large uncertainties in the formulation of QCD kinetic theory and in
the validity of algorithms employed to simulate off-shell propagation,
screening effects, color coherence, parton production and absorption, and
hadronization, it is important to develop and subject codes to standardized
tests on problems where analytic solutions are known. One reason for
concern is that cascade models necessarily violate Lorentz invariance through
the scattering prescriptions involving action at a distance
\cite{sorg0,noack,kodama,pratt,zhangpang97}. At RHIC and LHC energies, the
phase space of partons span over ten units of rapidity and extreme Lorentz
contraction effects arise which amplify non-causal numerical artifacts.
Numerical prescriptions have been proposed to minimize such 
Lorentz violations effects\cite{zhangpang97}. 
However, without further tests it is impossible
to ascertain whether  the final numerical
results are in fact physically correct.  Different
Monte-Carlo event generators can give  different predictions even for
basic observables such as the transverse energy per unit rapidity,
$dE_\perp/dy$.

In this paper we propose a simple test of the multiple collision algorithm of
such models through the computation of the evolution of $dE_\perp/dy$ in the
special case of longitudinal boost and transverse translation invariant thermal
initial conditions.  This observable is of interest because it 
is the simplest one
sensitive to collective hydrodynamic phenomena.  It is known\cite{Ruus1} that
for such boundary conditions, Euler hydrodynamics predicts that the final
observed transverse energy is reduced by a factor of about two relative to the
initial value due to $pdV$ work associated with longitudinal
expansion\cite{gyumat84}. This very basic feature of ultra-relativistic
nuclear 
reactions arises even if the initial conditions are highly inhomogeneous and
turbulent as shown recently 
in \cite{Gyul2}. On the other hand, pQCD estimates suggest
that the viscosity of a quark gluon plasma is very high, and
Navier-Stokes hydrodynamic equations predict about a factor of two
 less work is done by the plasma. The observable consequence
of such dissipative dynamics is  the reduction
of the transverse energy loss predicted by Euler
hydrodynamics\cite{gyudan,Esko1}.  
In fact,the parton mean free paths may be so long that 
 even
the Navier-Stokes approximation.  It
appears essential therefore to employ a full microscopic transport theory to
evaluate precisely the magnitude of that transverse energy loss. The newly
developed parton cascade models provide an approximate numerical solution to
the underlying covariant transport theory and are thus ideally suited to address
this problem.

In \cite{baym84,kajmat85,gavin91} analytic solutions 
for the evolution of the energy density were found
using  relativistic Boltzmann kinetic theory\cite{degroot}.
In this paper we extend those solutions to initial conditions
more relevant to the RHIC energy domain and propose these
to test parton cascade models.

The discussion is organized as follows:
In section 2, we review essential elements of longitudinally
boost invariant kinetic theory. The free streaming solution 
is discussed for both the ideal inside-outside and local
thermal initial conditions. The integral equation for the phase
space distribution in the relaxation time
approximation is presented. In section 3, the integral equation
for the energy density is derived together with its
Navier-Stokes limit for both constant and scaling relaxation times.
In section 4, the integral equation for the
transverse energy evolution is derived and solved numerically.
The results show that for the expected initial conditions
based on the HIJING event generator\cite{Wangmg},
kinetic theory  leads to much less collective transverse
energy loss than Euler hydrodynamics but more than Navier-Stokes.
We show that at least one of the newly developed models,
ZPC\cite{zhangpang97}, is able to  pass this test. Finally, we show that
the ZPC also reproduces the kinetic theory solution under more extreme
conditions where the kinetic theory solution approaches the Navier-Stokes
solution. As far as we know, this is the first demonstration of a parton
cascade solution approaching an analytic Navier-Stokes result.
Physically, this result is of interest because it shows that in order to
approach
the Navier-Stokes regime initial parton densities must be at least a factor of
four greater than predicted by the HIJING event generator\cite{Wangmg} 
for central $Au+Au$
collisions at RHIC.

\section{Boost Invariant Relativistic Transport}

The covariant  kinetic equation 
 for the invariant on-shell phase space density, $f(x,p)$ is 
\beq (p\partial_x)
f(x,p)= (pu) (C(x,p) + S(x,p) )\eeq{km1} 
where $C$ is the Boltzmann collision
term, $S$ is the source term, and $u^\mu$ is the collective
flow velocity field.  This equation greatly simplifies
in the case that longitudinal boost invariance and
transverse translation invariance is
assumed\cite{baym84,kajmat85,gavin91}. In that case, the on-shell
phase space distribution function depends  only on the reduced
phase space variables
$(\tau,\xi,\vpp)$, where $\tau^2=t^2-z^2$ and $\xi=\eta-y$ in terms of the
kinetic rapidity $y=\tanh^{-1}(p^z/p^0)$ and pseudo-rapidity
$\eta=\tanh^{-1}(z/t)$.  The collective
flow velocity field in this case
is  $u^\mu=x^\mu/\tau=(\ch\eta,0_\perp,\sh\eta)$ with
$(pu)=m_\perp\ch\xi$ and the transport equation reduces to 
\beqar (p\partial_x)f(x,p)
&=&m_\perp \left(\ch(\xi)\frac{\partial}{\partial\tau} -
\frac{\sh(\xi)}{\tau} \frac{\partial}{\partial\xi}\right)
f(\tau,\xi,\vpp)\nonumber\\ 
&=&(pu)\frac{d}{d\tau} f(\tau,\bar{\xi}(\tau),\vpp) 
\; \; .
\eeqar{km2} 
The
characteristic function $\bar{\xi}(\tau)\equiv\bar{\xi}(\tau;\tau_0,\xi)$
 satisfies
\beq
d\bar{\xi}/d\tau= -\tau^{-1}\tanh\bar{\xi}
\; \; ,\eeq{xi} 
and passes through $\xi=\bar{\xi}(\tau_0;\tau_0,\xi)$
at  $\tau=\tau_0$.
This trajectory  satisfies
 \beqar
t\;\sh\bar{\xi}(t;\tau,\xi) &=&\tau\;\sh\xi \nonumber \\
\ch\bar{\xi}(t;\tau,\xi)&=& \left( 1+ (\tau/t)^2\sh^2\xi\right)^{1/2} 
\; \; . \eeqar{xi2}
We will drop  the $\tau$ or $\xi$ arguments on $\bar{\xi}(t)$
when confusion cannot arise.

In
\cite{kajmat85} the source of partons was modeled by a Schwinger pair
production. That mechanism  models
soft beam
jet fragmentation for lab energies $<200$ AGeV.
At ultra-relativistic collider energies $\surd{s}>100$ AGeV, 
 perturbative QCD minijet production is currently thought to dominate
the parton production mechanism
\cite{Wangmg,Esko1}.  Given an initial distribution
of massless partons, e.g. from HIJING\cite{Wangmg}, 
\beq
g(y,\vpp)=dN_g/dyd^2 p_\perp
\; \; , \eeq{g}
the first problem 
is to construct the source  phase space distribution, $S(x,p)$. 
\subsection{Free Streaming Case}

To construct the source, consider an ensemble of on-shell 
partons with production phase space
coordinates, $\{x^\mu_a,p^\mu_a;p_a^2=m^2\}$.
The general form of the covariant free streaming phase space distribution
is given by \cite{degroot}
\beqar
{\cal F}(x,p) &=& \int ds 
\langle \sum_a P(s,p_a) \delta^4(x^\mu-x^\mu_a-p^\mu_a s)
\delta^4(p-p_a)\rangle
\;\; ,
\eeqar{f00}
where $P(s,p)$ is introduced  as the   Lorentz  scalar 
probability that a parton with four momentum $p$
has been formed by proper time $\tau=s\;m$. This of course
depends
on details of the formation dynamics and  external four vectors
 and parameters specifying the reaction.

The  on-shell constraint leads to 
\beqar
\delta^4(p-p_a)&=& 2\theta(p_0)\delta(p^2-m^2) \delta(y-y_a)\delta^2(\vpp-
\vp_{\perp a})
\; \; , \eeqar{del1}
and thus the  reduced (6+1) dimensional phase space density, $ {f}$,
is defined through
\beq
{\cal F}(x,p)=2 \theta(p^0)\delta(p^2-m^2)  {f}(x,p)
\; \; . \eeq{fred}
The on-shell phase space source function in eq.(\ref{km1})
is then given by 
\beq
S(x,p)= \frac{1}{(pu)}(p\partial_x) f(x,p) 
\; \; . \eeq{source} 
For boost invariant boundary conditions,
\beq
 {S}(\tau,\xi,\vpp)=\frac{d}{d\tau}
 {f}(\tau,\bar{\xi}(\tau,\tau^\prime,\xi)),\vpp)|_{\tau^\prime=\tau}
\; \; . \eeq{tilds}

The local 
energy-momentum tensor and parton current are  obtained as
\cite{degroot}
\beqar
T^{\mu\nu}(x)&=&\int d^4p \; p^\mu p^\nu {\cal F}(x,p)= 
\int \frac{d^3\vp}{p^0} \; 
p^\mu p^\nu  {f}(x,p) \nonumber \\
J^\mu(x)&=& \int d^4p \; p^\mu {\cal F}(x,p)=\int \frac{d^3\vp}{p^0} \; 
p^\mu   {f}(x,p)
\; \; . \eeqar{deftj}
The proper energy density and number density are in turn given by
$\epsilon(x)= u_\mu u_\nu T^{\mu\nu}(x),
\; \rho(x)=u_\mu J^\mu(x)$
We use here the 
Landau definition \cite{degroot} of the normalized fluid four 
velocity $u^\mu(x)=T^{\mu\nu}u_\nu/(uTu)$, such
that the energy flux,
$T^{0i}$, vanishes in the comoving frame where $u^\mu=(1,0,0,0)$.

The central observable  of interest is
the parton  invariant momentum distribution 
 as given by the Cooper-Frye formula
\beq
E\frac{d^3 N}{d^3 p}= \int_\Sigma d\Sigma_\mu p^\mu  {f}(x,p)
\eeq{cf}
Here  $\Sigma^\mu(\zeta_1,\zeta_2,\zeta_3)$ specifies the 3D hypersurface
on which the momentum distribution is measured. The
volume element is given by
$$d\Sigma_\mu = \epsilon_{\mu\alpha\beta\gamma}
 (\partial \Sigma^\alpha/\partial\zeta_1)
 (\partial \Sigma^\beta/\partial\zeta_2)
 (\partial \Sigma^\gamma/\partial\zeta_3) d\zeta_1 d\zeta_2
d\zeta_3 \; \; .$$
For a fixed time measurement, we  take $(\zeta_1,\zeta_2,\zeta_3)=(\vxp,z)$,
$\Sigma^\mu=x^\mu=(t,z,\vxp)$ 
so that  $d\Sigma_\mu=(d^3x,{\bf 0})$.
The inclusive parton distribution at time $t$ is then given by
\beqar
E\frac{d^3 N}{d^3 p}(t)&=& \frac{dN(t)}{dyd^2\vpp}=
E \int d^3x  {f}(x,p)
\; \; ,\eeqar{cf1}
and the transverse energy per unit rapidity at time $t$ is
\beqar
\frac{dE_\perp(t)}{dy}&=&\int d^2 \vpp \; m_\perp \frac{dN}{dyd^2 \vpp}
\; \; .\eeqar{det1}

On the other hand, for the physically more relevant
longitudinal boost invariant boundary conditions, 
 the final parton distribution requires an integration over
 a  fixed proper time  freeze-out surface.
 For 
$\tau=(t^2-z^2)^{1/2}=\tau_f$, $\Sigma$ is conveniently parameterized by 
$\Sigma^\mu=(\tau_f\ch\eta,\tau_f\sh\eta,\vxp)$
with $d\Sigma^\mu=(\tau_f\ch\eta,-\tau_f\sh\eta,0)d\eta d\vxp$.
Therefore, $d\Sigma^\mu p_\mu=(\tau_f m_\perp\ch\xi)d\eta d\vxp$.
In this case,
\beqar
\frac{dN(\tau_f)}{dyd^2\vpp}=
 \int \tau_f d\eta d^2\vxp \;m_\perp \ch\xi \;  {f}(x,p)
\; \; , \eeqar{cf2}
and the basic calorimetric observable is
\beqar
\ep(\tau_f)&\equiv &\frac{dE_\perp(\tau_f)}{dy}
=\int \tau_f d\eta d^2\vxp\int d^2\vpp m_\perp^2\ch \xi\;  {f}(x,p)
\eeqar{det2}
Our main focus here is on how $\ep(\tau)$  evolves
to its final form given an initial boost invariant value 
at $\tau=\tau_0$.

\subsection{Ideal Inside-Outside Correlation}

In  nuclear collisions at ultra-relativistic energies,
the initial parton coordinates
are approximately localized on a transverse sheet,
with 
$x^\mu_a=(0,\vx_{\perp a},0)$ with $\vx_{\perp a}$ 
distributed over a  transverse 
area according to a density $\rho(\vx_\perp)\approx
\theta(R-r)/\pi R^2$. Transforming to the $(\tau,\xi=\eta-y,\vpp)$ coordinates
\beqar
\delta^4(x-x_a-p s)
&=&  \frac{\delta(\xi)}{m_\perp\tau} \delta(s-\tau /m_\perp)
\delta^2(\vxp-\vx_{\perp a}-\frac{\vpp}{m_\perp} \tau)
\; \; .\eeqar{del2}
Neglecting transverse expansion, the approximate
ideal inside-outside correlated distribution is given by
\beqar
f_0(\tau,\xi,p_\perp)
&\approx &  \frac{\delta(\xi)}{m_\perp
\tau\pi R^2} 
 P(\tau m_\perp)
 g(\vpp)
\eeqar{f0}

The influence of the formation physics
can be studied by comparing  two analytic models of $P=P(m_\perp\tau)$:
\beq
P( m_\perp\tau )=\left\{
\begin{array}{ll}
\theta(\tau-\tau_0) & Model \; I \\
( m_\perp\tau)^2/(1+( m_\perp\tau)^2)& Model \; II
\end{array} \right.
\eeq{lorentz}
Model I is simplest for analytic tests, while
Model II is more realistic based
on  a study \cite{Gyul1} of induced radiation in multiple collisions.
This is obviously one of the most uncertain aspects of parton cascade 
models since formation physics is beyond the scope of ballistic kinetic
 theory. At the least, kinetic theory could be supplemented
by classical field evolution such as proposed in\cite{mcler}.
(See \cite{Esko1} for example for solutions to  coupled field and kinetic
equations in the chromo-hydrodynamics limit.)

The phase space source function given (\ref{f0})  is 
\beq
 {S}_0(x,p)=\frac{d}{d\tau}
\frac{\delta(\bar{\xi}(\tau)){P}( m_\perp\tau) }{m_\perp\tau\pi R^2} 
 g(\vpp)=\frac{\delta(\xi)\dot{P}( m_\perp\tau) }{\tau\pi R^2}
g(\vpp)
\; \; , \eeq{source0}
where we used 
\beq
\delta(\bar{\xi}(t;\tau,\xi))=\frac{t}{\tau}\delta(\xi)
\; \; . \eeq{delxi} 
For this ideal $\xi=0$ correlated case, the energy-momentum tensor 
has the form
$T_0^{\mu\nu}(x)
u^\mu u^\nu \epsilon_0(\tau) + \delta^{\mu\nu}_\perp P_T(\tau)
$
where $\delta^{\mu\nu}_T={\rm diag}(0,0,1,1)$.
The proper energy density evolves as
\beq
\epsilon_0(\tau)= (uTu)
=\frac{\langle m_\perp P(m_\perp \tau)\rangle}{
\tau \pi R^2}
\; \; , \eeq{eps0}
where the brackets are used to indicate the transverse momentum integration
$\langle \cdots\rangle \equiv \int d^2 \vpp \; g(\vpp)\cdots $
Note that for the ideal correlated case, 
 the proper longitudinal pressure vanishes,
while the transverse pressure, 
$P_T=\epsilon/2$, is $50\%$ larger than
in thermal equilibrium for massless partons. 

The hydrodynamic equations follow from 
$((\partial T) u)=\sigma$ where
\beq
\sigma(\tau)=
\int d\xi d^2\vpp m^2_\perp \ch\xi^2 S(\tau,\xi,\vpp) 
\eeq{ptmn}
 is the proper energy density source per unit time.
The proper energy density  evolves as
\beq
\dot{\epsilon} +\frac{\epsilon}{\tau}= 
\de(\tau)
=\frac{\langle
m_\perp^2 \dot{P}(m_\perp \tau)\rangle}{\tau \pi R^2} 
\eeq{bj}
The well known solution is  given by $\epsilon(\tau)=\epsilon_0(\tau)$ 
from 
eq.(\ref{eps0}).
For the Lorentzian (model II) formation probability in
eq.(\ref{lorentz}), $\epsilon$ 
increases initially linearly with proper time until
$\tau_0\sim 1/\langle m_\perp\rangle$ and then eventually decrease
as $1/\tau$ due to one dimensional longitudinal expansion.

The inclusive distribution (\ref{cf1}) evolves in this case as
\beqar
\frac{dN(t)}{dyd^2\vpp}&=& \ch(y) \int dz 
\frac{\delta(\eta-y)}{\tau} P(m_\perp\tau) g(\vpp)
\nonumber \\
&=& 
g(\vpp) P\left(\frac{m_\perp t}{\ch(y)}\right) 
\eeqar{cf0}
showing  that partons
are formed   according to a  time dilated formation probability.
The transverse energy (\ref{det1}) develops as
\beqar
\frac{dE_\perp(t)}{dy}
&=& \langle m_\perp P(m_\perp t/\ch(y))\rangle 
\eeqar{det0}
On the other hand, along a freeze-out  
  proper time surface 
$\tau=\tau_f$, the inclusive distribution (\ref{cf2}) is given by
\beqar
\frac{dN(\tau_f)}{dyd^2\vpp}
=g(\vpp) P(m_\perp \tau_f) 
\; \; . \eeqar{cf22}
Thus $g$ is the asymptotic
invariant parton distribution in the free streaming case.
The final transverse energy per unit rapidity (\ref{det2}) is
\beqar
\ep(\tau_f)&\equiv &\frac{dE_\perp(\tau_f)}{dy}
=\int \tau_f d\eta d^2\vxp\int d^2\vpp m_\perp^2\ch \xi\;  {f}(x,p)
\nonumber \\
&=& \langle m_\perp P(m_\perp \tau_f) \rangle 
= (\tau_f \pi R^2)\; \epsilon_0(\tau_f) 
\eeqar{det22}
On account of (\ref{bj}), $dE_\perp/dy$
tends toward  a constant after the formation
era. The last line is just the familiar Bjorken formula
relating the observable $dE_\perp/dy$ to the energy density
at the freeze-out proper time $\tau_f$. 

\subsubsection{Free Streaming  with Thermal Initial Conditions}

In order to study the influence of 
more realistic  finite width of the $\eta-y$ initial state correlations, 
it is instructive
to start with  local thermalized source 
\beq
 {S}_{th}(\tau,\xi,p_\perp)= \delta(\tau-\tau_0) 
 F\left(\frac{p_\perp\ch\xi}{T_0}\right)
\eeq{sth}
where, eg., $F(x)=ce^{-x}$.
For such a  source, the solution of the free streaming transport
equation is 
\beq
 {f}_{1}(\tau,\bar{\xi}(\tau;\tau_0,\xi),p_\perp)=\theta(\tau-\tau_0)
 F\left(\frac{p_\perp\ch\xi}{T_0}\right)
\; \; , \eeq{fth} 
and therefore inverting with (\ref{xi2}),
\beq
 {f}_{1}(\tau,\xi,p_\perp)=\theta(\tau-\tau_0)
F\left(\frac{p_\perp}{T_0} \ch\bar{\xi}(\tau_0;\tau,\xi)
\right)
\; \; , \eeq{fth1} 
where
\beq
\ch\bar{\xi}(\tau_0;\tau,\xi) =\left(1+(\tau/\tau_0)^2 \sh^2\xi\right)^{1/2}
\; \; . \eeq{chbarxi}
The free streaming energy density in this case is 
\beqar
\epsilon_1(\tau)&=&
\int d\xi d^2\vpp \; p_\perp^2\ch^2\xi  {f}_1(\tau,\xi,
p_\perp)= 
\theta(\tau-\tau_0)\epsilon(\tau_0) \frac{\tau_0}{\tau} h\left(\tau_0/\tau
\right)
%
%
\; \; , \eeqar{bj2}
where  the function $h$ modulating the Bjorken $\tau_0/\tau$ factor
is the same as the one defined in  \cite{kajmat85,gavin91}:
\beqar
x h(x)&\equiv&w(x)=x^4 \int_{-\infty}^\infty \frac{d\xi}{2}\frac{\ch^2\xi}{
(x^2+\sh^2\xi)^2}
=\frac{x}{2}\left(x+\frac{\sin^{-1}\sqrt{1-x^2}}{
\sqrt{1-x^2}}\right)
 \nonumber \\
&\approx& 1 - \frac{4}{3}(1-x) + \frac{2}{5}(1-x)^2 +\cdots 
\; \; . \eeqar{xi3}
For large times, $x\rightarrow 0$ and 
\beq
xh(x)\rightarrow \frac{\pi}{4} x
\; \; . \eeq{xi30}
The factor $\pi/4$ is characteristic of the difference between
the isotropic and inside-outside correlated phase space.
 The energy density at late times
follows Bjorken formula reduced by this factor, $\epsilon(\tau)
\approx \frac{\pi}{4} \epsilon(\tau_0) (\tau_0/\tau)$.
The initial local thermal energy density is given by
\beq \epsilon(\tau_0)=\int d\xi d^2\vpp \; p_\perp^2\ch^2\xi  
F(p_\perp\ch\xi/T_0) = K_{SB}T_0^4
\; \; ,\eeq{th0}
where the Stefan-Boltzmann constant is
\beq 
K_{SB}=4\pi \int_0^\infty dx x^3 F(x)
\; \; .\eeq{k}

The transverse energy per unit rapidity is closely 
related to the energy density as seen in eq.(\ref{det2}).
\beqar
\ep^1(\tau)&=&
(\tau \pi R^2)\int  d\eta  d^2\vpp p_\perp^2\ch \xi\; 
 {f}_1(\tau,\xi,\vpp)
\; \; . \eeqar{et1}
This differs from $\epsilon_1(\tau)$ in eq.(\ref{bj2}) by only one power less
of
$\ch\xi$ in the integrand.
Noting  that
\beqar
\int_{-\infty}^\infty \frac{d\xi \ch\xi}{\ch^4\bar{\xi}(t;\tau,\xi)}
=\frac{\pi}{4}\frac{t}{\tau}
\; \; , \eeqar{w3}
 the transverse energy per unit rapidity
remains a constant during free streaming
\beqar
\ep^1(\tau)&=&
 \frac{\pi}{4} (\tau_0 \pi R^2) \epsilon(\tau_0)=\lim_{\tau 
\rightarrow\infty} (\tau\pi R^2)\epsilon_1(\tau)
\; \; , \eeqar{eteq1}
unlike the energy density.
The Bjorken relation between $\ep$ and $\epsilon$ is 
recovered in this
 case only at large times, after the initial local isotropic 
conditions evolve 
toward the ideal $\xi=0$ case.

\subsection{Collisions in the Relaxation Time Approximation}

The simplest way to extend the free streaming 
analysis to include the effects
of collisions in eq.(\ref{km1})
is via a relaxation time approximation\cite{baym84,kajmat85}:
\beq
C(x,p)=-(pu)(f-f_{eq})/\tau_c(x)
\eeq{coll}
where the equilibrium distribution, 
$f_{eq}(x,p)$
is constrained
by  energy momentum conservation $(\partial T u=0)$
to obey
\beq
\int \frac{d^3 \vp}{p^0}  p^\mu (pu) (f-f_{eq})/\tau_c =0
\; \; . \eeq{const} 
While the relaxation time approximation is strictly valid
only for 
small deviations from local equilibrium,
it  provides one  important limit where the general
nonlinear transport
equations encoded in parton cascade models
can be subjected to an analytic test.

With longitudinal boost invariant boundary
conditions the equilibrium distribution depends only on
$(pu)$ and the local temperature $T(\tau)$,
\beq
 {f}_{eq}(x,p)= F\left(\frac{pu}{T(\tau)}\right)
= {f}_{eq}\left(
\frac{m_\perp \ch\xi}{T(\tau)} \right)
\;\; .
\eeq{therm}
Eq. (\ref{const}) then constrains the temperature to evolve as
\beq
\epsilon(\tau)=\epsilon_{eq}(\tau)=   K_{SB}T^4(\tau)
\eeq{equil}
as long as $\tau_c$ is assumed to be independent of $p$.

The transport equation along 
the  characteristic $\bar{\xi}(t;\tau,\xi)$
that passes through $\xi$ at proper time $\tau$ is
\beq
\frac{d}{d t} {f}(t,\bar{\xi}(t;\tau,\xi),\vpp) =-
\gamma_c(\tau) ({f}-{f}_{eq})(t,\bar{\xi}(t;\tau,\xi),\vpp) + {S}(
t,\bar{\xi}(t;\tau,\xi) ,\vpp)
\; \; .\eeq{til1}
This is solved\cite{kajmat85} 
using $e^{-\beta} \partial_{\tau} (e^{\beta} f)$
to shift $\gamma_c {f}$ to the left hand side
of the equation taking 
\beq
\beta_c(\tau)=\int_0^\tau dt \gamma_c(t)
\; \; ,\eeq{beta}
in terms of the  collision rate
\beq
\gamma_c(\tau)=\tau_c^{-1}(\tau)=\dot{\beta}_c(\tau)
\eeq{gam}
The collision rate is assumed to vanishes at $\tau=0$
for our boundary conditions.

The integral equation for the phase space density is therefore
\beq
e^{\beta_c(\tau)}{f}(\tau,\xi,\vpp)
= \int_0^\tau dt e^{+\beta_c(t)}\left(\dot{\beta}_c(t)
{f}_{eq}(t,\bar{\xi}(t;\tau,\xi),p_\perp)\;+\; {S}(t,\bar{\xi}(t;\tau,\xi),\vpp) \right)
\; \; .\eeq{til2}
Defining the scattering kernel $K(t,\tau)$ via
\beq
K(t,\tau)= \dot{\beta}_c(t) e^{\beta_c(t)-\beta_c(\tau)}  =
\gamma_c(t)
\exp\left(-\int_t^\tau \frac{dt^\prime}{\tau_c(t^\prime)}
\right)
\; \; ,\eeq{kern}
 the integral equation for $f$ can be written as follows:
\beq
{f}(\tau,\xi,\vpp)= \int_0^\tau dt K(t,\tau) \left(
{f}_{eq}(t,\bar{\xi}(t;\tau),p_\perp)
+\tau_c(t) S(t,\bar{\xi}(t;\tau),p_\perp) \right)
\; \; . \eeq{til22}
Note that the kernel  tends toward $\delta(t-\tau)$ 
in the  rapid thermalization limit, $\tau_c(\tau)\rightarrow 0$,
and that in that case $f\rightarrow f_{eq}$ in the source free region.

For the special case of boost invariant thermal initial
conditions, eq.(\ref{sth}), $f$ evolves for $\tau>\tau_0$ 
according to 
\beqar
{f}(\tau,\xi,\vpp)&=&\theta(\tau-\tau_o) e^{\beta_c(\tau_0)-\beta_c(\tau)}
F(p_\perp\ch\bar{\xi}(\tau_0;
\tau)/T_0)\\ \nonumber 
&\;& \; \;\;\;\;
+ \int_{\tau_0}^\tau dt K(t,\tau) 
{f}_{eq}(p_\perp \ch\bar{\xi}(t;\tau)/{T(t)})
\; \; .\eeqar{til21}
This also shows how the system forgets its initial condition as it 
tries to evolve toward  local thermal equilibrium. Whether it
get close to $f_{eq}$ or how far it lags behind depends of
course on the functional form of the collision rate $\gamma_c(\tau)$.

\section{Evolution of the Energy Density}

The integral equation 
for the proper energy density 
is obtained by integrating eq.(\ref{til22}) using
\beq
\epsilon(\tau)= \int d\xi d^2\vpp p_\perp^2 \ch^2\xi\;
 {f}(\tau,\xi,\vpp)
\; \; . \eeq{ep2}
This leads to
\beqar
\epsilon(\tau)&=& 
\int_0^\tau dt K(t,\tau)\int d\xi d^2\vpp p_\perp^2 \ch^2\xi\;
\left( 
{f}_{eq}(t,\bar{\xi}(t;\tau),p_\perp) +\tau_c(t)S(t,\bar{\xi}(t;\tau)
,p_\perp) 
\right)\nonumber\\
&\;&
\eeqar{eps1}
The transverse momentum and $\xi$ integration over $f_{eq}$ can
be done as in section2.3 yielding
a factor $K_{SB}T^4(t) w(t/\tau)$. 
The main simplification arises as a result of
the constraint  (\ref{equil}) that allows us to replace
$K_{SB}T^4(t)$ by $\epsilon(t)$.
Therefore,
\beqar
\epsilon(\tau)&=& 
\int_0^\tau dt K(t,\tau)\left( \epsilon(t)w(t/\tau)
+\tau_c(t)\bar{\sigma}(t) 
\right)
\; \; , \eeqar{epsf}
where the  source term 
depends on the form of the formation 
$\xi$ correlations
via
\beqar
\bar{\sigma}(t)= \int d\xi d^2\vpp p_\perp^2 \ch^2\xi\;
 {S}(t,\bar{\xi}(t;\tau,\xi),\vpp)
\; \; . \eeqar{sigbar}
For ideal $\xi=0$ correlations, $\bar{\sigma}$, reduces to
eq.(\ref{ptmn}). For a thermal correlated source as in, eq.(\ref{sth}), 
 $\bar{\sigma}$
reduces to $\sigma(t)$ modulated by the same $w(t/\tau)$
factor, (\ref{xi3}), as the $\epsilon(t)$ term in (\ref{epsf}):
\beqar
\epsilon(\tau)&=& 
\int_0^\tau dt K(t,\tau)w(t/\tau)\left( \epsilon(t)
+\tau_c(t){\sigma}(t) 
\right)
\nonumber\\
&=& \theta(\tau-\tau_0)\epsilon(\tau_0)\;
\tau_c(\tau_0)K(\tau_0,\tau)\; w(\tau_0/\tau) \nonumber\\
&\;& \; \; \;\;\; 
+ \int_0^\tau dt K(t,\tau)w(t/\tau)\epsilon(t)
\; \; . \eeqar{epsfth}
In the free streaming limit, $\tau_c\rightarrow \infty$, $K\rightarrow 0$
but $\tau_c K\rightarrow 1$. In that limit $\epsilon$ reduces to $\epsilon_1$
in eq.(\ref{bj2}). In the opposite, rapid thermalization limit
(\ref{epsfth}) reduces to hydrodynamics as we show in the next section.

\subsection{Navier-Stokes Limit: Constant \protect{$\tau_c$}}

In the limit of rapid thermalization, $\tau_c\rightarrow 0$
 we can expand  eq.(\ref{epsfth}) in a power series in  $\tau_c$.
Consider first the  case where  $\tau_c$ independent
of time\cite{baym84,kajmat85}. In this case 
\beq
K(t,\tau)=e^{(t-\tau)/\tau_c}/\tau_c
\; \; . \eeq{ker1}
The kernel is highly peaked at $t=\tau$. Thus for slowly varying
functions, $F(t)$ we can systematically expand
\beqar
\int_0^\tau dt \;K(t,\tau) F(t)&=&\int_0^{\tau/\tau_c}
dx e^{-x}(F(\tau)-\tau_c x\dot{F}(\tau)+\frac{1}{2}
 \tau_c^2 x^2 \ddot{F}(\tau)
+\cdots \nonumber\\
&=& F(\tau)-\tau_c\dot{F}(\tau)+\tau_c^2 \ddot{F}(\tau)+\cdots
\eeqar{expan1}
The formal expansion of (\ref{epsfth}) to second order in $\tau_c$
then leads to
\beqar
\epsilon(\tau)&=&\epsilon(\tau)-\tau_c(\dot{\epsilon}+\frac{4}{3}
\frac{\epsilon}{\tau}-\sigma(\tau) )\\\nonumber
&\;& \;+\tau_c^2(\ddot{\epsilon}+\frac{8}{3}\frac{\dot{\epsilon}}{\tau}
+\frac{4}{5}\frac{\epsilon}{\tau^2} -\dot{\sigma}-\frac{4}{3}
\frac{\sigma}{\tau})
\; \; , \eeqar{exp12}
where  we  $w(1)=1,
\dot{w}(1)=4/3,\ddot{w}(1)=4/5$ has been used.

The full integral equation eq.(\ref{epsfth}) reduces 
in this case to
\beqar
\left(\dot{\epsilon}+\frac{4}{3}\frac{\epsilon}{\tau} -
\sigma \right)&=&
\tau_c\left( \ddot{\epsilon} +\frac{8}{3}\frac{\dot{\epsilon}}{\tau}+
\frac{4}{5}\frac{\epsilon}{\tau^2}
-\dot{\sigma}-\frac{4}{3}\frac{\sigma}{\tau}
\right)  + O(\tau_c^2) 
\eeqar{navst}
To lowest  order in $\tau_c$,  (\ref{navst}) reduces to the
Euler hydrodynamic equations for  Bjorken
boundary conditions
\beqar
\dot{\epsilon}&\approx& -\frac{4}{3}\frac{\epsilon}{\tau}+
\sigma
+O(\tau_c) 
\; \; .  \eeqar{euler}
Thus to lowest order
\beqar
\ddot{\epsilon}&\approx& \frac{4}{3}\left(\frac{4}{3}+1\right)
\frac{\epsilon}{\tau^2}+\dot{\sigma}-\frac{4}{3}\frac{\sigma}{\tau}
+O(\tau_c) 
\; \; .  \eeqar{deuler}
The first order correction term can therefore be simplified  using
\beqar
\ddot{\epsilon} + \frac{8}{3}\frac{\dot{\epsilon}}{\tau}+
\frac{4}{5}\frac{\epsilon}{\tau^2}
=\frac{4}{3}\frac{4}{15}\frac{\epsilon}{\tau^2} + 
 \dot{\sigma} + \frac{4}{3}\frac{\sigma}{\tau} 
+ O(\tau_c)
\; \; . \eeqar{exp4}
Note that the $\dot{\sigma}$  terms in (\ref{navst}) cancel,
and eq.(\ref{navst}) reduces to 
the Navier-Stokes, viscous hydrodynamic equation
\beq
\dot{\epsilon}+\frac{4}{3}\frac{\epsilon}{\tau}=
\frac{4}{3}\frac{\eta}{\tau^2} \; + \;
\sigma(\tau) 
\;+\;O(\tau_c^2) 
\; \; , \eeq{nsf}
with the expected\cite{degroot,gyudan}
 shear viscosity coefficient  given by
\beq
\eta=\frac{4}{15} \tau_c \epsilon
\; \; . \eeq{visc}
For constant $\tau_c$ it is clear that viscous corrections can be neglected
at large times.

\subsection{Navier-Stokes in the Scaling Limit: {$\tau_c=\tau/\alpha$}}

For the Bjorken boundary conditions, however,
the proper density $\rho=(uJ)$ decreases
as $\rho(\tau)=\tau_0\rho(\tau_0)/\tau$ due to longitudinal expansion.
Consequently, the collision rate is expected to scale\cite{gavin91}
as
\beq
\gamma_c=\tau_c^{-1}=
\langle\sigma_t \rho(\tau)\rangle\equiv \alpha/\tau
\eeq{scalt}
The  coupling parameter that controls the
rate of thermalization is in this case approximately constant
\beq
\alpha=\langle\sigma_t\rho(\tau_0)\tau_0\rangle=\tau_0/\tau_c(\tau_0)
\eeq{alf}
where $\sigma_t$ is the transport cross section,
and $\tau_0$ is any time after the source is
negligible. For a  finite formation probability, such as in eq.(\ref{lorentz}),
$\alpha$ must initially grow from zero.
In this section we ignore this initial time dependence
of $\alpha$ to simplify the analytic treatment.
In this scaling regime, 
the kernel is
\beq
K(t,\tau)= \frac{\alpha}{t}
\left(\frac{t}{\tau}\right)^{\alpha}
\eeq{ker2}
The integral equation (\ref{epsfth}) for the energy density reduces 
 to \cite{gavin91}
\beq
\epsilon(\tau)= 
\alpha \int_{0}^\tau \frac{dt}{t}
\left(\frac{t}{\tau}\right)^{\alpha+1}h(t/\tau)(
 \epsilon(t) + t\de(t)/\alpha)
\eeq{epsf3}

For $\alpha\gg 1$ the kernel is strongly peaked near $t=\tau$,
and we can expand,  analogous to eq.(\ref{expan1}),  in powers of $1/\alpha$ 
\beqar
\int_0^\tau dt \;K(t,\tau) F(t)&=&\alpha \int_{0}^{1}
dx x^{\alpha-1} F(\tau-\tau(1-x))\nonumber \\
&\stackrel{\alpha\rightarrow \infty}{\longrightarrow}&
F(\tau)- \frac{\tau\dot{F}(\tau)}{\alpha+1}+\frac{\tau^2 \ddot{F}(\tau)}{
(\alpha+1)(\alpha+2)} 
+\cdots 
\nonumber \\
&=&
F(\tau)- \frac{\tau\dot{F}(\tau)}{\alpha}+\frac{\tau\dot{F}(\tau)
+\tau^2 \ddot{F}(\tau)}{
\alpha^2} 
+ O(\frac{1}{\alpha^3})\nonumber\\
\; \; .\eeqar{expan3}
The source term expanded to second order in this case is 
\beqar
\frac{\tau\de}{\alpha}- \frac{\tau^2\dde+\frac{7}{3}\tau\de}{\alpha^2}
+O(1/\alpha^3)
\; \; ,  \eeqar{sou4}
while the energy density  term expands to 
\beqar
\epsilon-\frac{1}{\alpha}
(\tau\dot{\epsilon}+\frac{4}{3}{\epsilon}) 
+ \frac{1}{\alpha^2}
 \left(\tau^2 \ddot{\epsilon}+\frac{11}{3}\tau\dot{\epsilon}+
\frac{32}{15}\epsilon
 \right)
+O(\frac{1}{\alpha^3})
\; \; . \eeqar{exp5}
At  $O(1/\alpha)$, we again recover the Euler hydrodynamic equation
(\ref{euler}). To next order we recover
the Navier-Stokes equation (\ref{nsf})
with a time dependent viscosity
\beq
\eta(\tau)=\frac{4}{15}\frac{\tau\epsilon}{\alpha}
=\frac{4}{15}\tau_c(\tau)\epsilon(\tau)
\; \; . \eeq{visc2}
This has the same form as in (\ref{visc}) except that $\tau_c$ is here time
dependent.

The most interesting point associated with the scaling Navier-Stokes
is that unlike in the
constant $\tau_c$ case (\ref{nsf}), the viscosity term does not become
negligible compared to the pressure term ( $\frac{1}{3}\epsilon/\tau$)
at late times. Dissipation in this case decreases
the effective 
speed of sound from $c_s^2=1/3$, appropriate for an ideal 
ultra-relativistic gas,
to
\beq
c_s^2=\frac{1}{3}\left(1 -\frac{16}{15\alpha}\right)
\; \; . \eeq{cs}
As discussed in  \cite{gavin91}, this implies
that the energy density in the source free region
always decreases more slowly than Euler hydrodynamics predicts:
\beq
\epsilon(\tau)=\epsilon(\tau_0)
\left( \frac{\tau_0}{\tau} \right)^{\frac{4}{3}-\frac{16}{45\alpha}}
\; \; . \eeq{bjp}

It is important to emphasize however,
that the  Navier-Stokes approximation can only apply
if the viscous term is small compared to the pressure term.
This requires
\beq
\alpha\approx \frac{\sigma}{\pi R^2}\frac{dN}{dy} > 1
\;\;. \eeq{alfc}
This condition is certainly violated at early times when $dN/dy$ is 
small. In addition, as we will see, this condition is
not satisfied in a dense parton gas if the screened cross
sections are as small as pQCD estimates would indicate.
The detailed study of deviations of kinetic theory
and parton cascade solutions from  Navier-Stokes solutions 
is useful as a gauge of whether particular observables
such as transverse energy can be correctly 
interpreted in terms of thermodynamic
concepts.

\section{Evolution of \protect{$\ep(\tau)=dE_\perp/dy$}}

The transverse energy per unit rapidity is 
related to the phase space density via eq.(\ref{et1}).
For the case of local equilibrium, 
\beqar
\ep^{eq}(\tau)=
\int  \tau d\xi d^2\vxp\int  d^2\vpp p_\perp^2\ch \xi\; 
 {f}_{eq}(p_\perp \ch\xi/T(\tau)) 
= \frac{\pi}{4} (\tau \pi R^2) \epsilon_{eq}(\tau)
\; \; .\eeqar{eteq}
The integral equation for $\ep$ is analogous to (\ref{eps1})
involving one power less of $\ch\xi$ in the integrand
\beqar
\ep(\tau)&=& \int_0^\tau dt K(t,\tau) \int d\eta d^2\vpp 
 (\tau\pi R^2) p_\perp^2\ch\xi \\ \nonumber
&\;&\;\;\;\; \left(  
 {f}_{eq}(t,\bar{\xi}(t/\tau,\xi),p_\perp) 
+
\tau_c(t)  {S}(t,\bar{\xi}(t/\tau,\xi),\vpp)\right)
\;\; .\eeqar{etgen}
Because of eq.(\ref{w3})
the integral over $f_{eq}$ gives $\int K(t,\tau) e_\perp^{eq}(t)$
without the weight function $w$ of eq.(\ref{xi3}).

For local thermally correlated
sources, in particular, the integral equation reduces to
\beqar
\ep(\tau)
&=& \theta(\tau-\tau_0)\ep(\tau_0)\; \tau_c(\tau_0)K(\tau_0,\tau)
+
\int_{\tau_0}^\tau dt K(t,\tau)\ep^{eq}(t) 
\; \; ,\eeqar{etth}
Unfortunately, we cannot replace $  \ep^{eq}(t)$ by $  \ep(t)$,
as in the case of the energy density evolution equation,
because energy conservation only  constrains $\epsilon=\epsilon_{eq}$.
Therefore, we must first 
solve  the energy density integral equation (\ref{epsfth})
to compute the temperature, $T(\tau)$, and then use that
solution to compute $\ep^{eq}(\tau)$ via (\ref{eteq}). 
Numerically, it is simplest solve (\ref{epsfth},\ref{etth})
together in the same loop iterating time steps and keeping track of prior
time steps in an updated array of $\epsilon(\tau_n)$
\beqar
\epsilon(\tau_n)
&=& \tau_c(\tau_0)K(\tau_0,\tau) w(\frac{\tau_0}{\tau_n}) 
\epsilon(\tau_0) +\sum_{i=0}^{n-1} 
\Delta \tau K(\tau_i,\tau_n) w(\frac{\tau_i}{\tau_n})
 \epsilon(\tau_i) \\
\ep(\tau_n)
&=&\frac{\pi}{4}\pi R^2 \left(\tau_c(\tau_0)K(\tau_0,\tau)
\tau_0 \epsilon(\tau_0) +
\sum_{i=0}^{n-1} \Delta \tau K(\tau_i,\tau_n)\tau_i \epsilon(\tau_i)
\right)
\; \; .\eeqar{etfor}
This system converges very rapidly even for arbitrary
time varying relaxation rates, $\gamma_c(\tau)$, not limited
to the scaling form, (\ref{scalt}). Because of its numerical simplicity
we still refer to the solutions of the above system as analytic
for tests of parton cascade codes.

\section{Illustrative Numerical Tests}

In this section we solve the above integral equations for
$\ep(\tau)$ and compare them to hydrodynamics and parton cascade
results for an initial longitudinally boost invariant
but thermally correlated parton gas
with $\ep(\tau_0=0.2\;{\rm fm})=484$ GeV and very high parton density
$\rho(\tau_0=0.2\;{\rm fm})=20/{\rm fm}^3$
 initial conditions
for RHIC energies as estimated from
 HIJING\cite{Wangmg} simulations for 
central $Au+Au$ collisions at $\surd s=200$ AGeV.

First we show in Fig. 1 the expected hydrodynamic evolution
in the case of Euler non-dissipative and Navier-Stokes approximations.
 Note that the Euler solution decreases by over a factor of two
as expected from previous studies. The transverse energy loss is however
reduced
as the transport cross section is reduced. For $\sigma_t< 2mb$
(corresponding to $\alpha<0.8$) the Navier-Stokes
approximation fails badly even for this very high initial parton density.
In the figure we have in fact replaced the unphysical
negative effective pressure in the  Navier-Stokes calculation by zero.

In Fig. 2 we compare the kinetic theory solutions of (\ref{etfor}) 
to Navier-Stokes.
For $\sigma=32$ mb ($\alpha=12.8$) 
the transport and Navier Stokes solutions are almost
identical. However,
by $\sigma=3$ mb, the Navier-Stokes approximation significantly under-predicts
the work done during the expansion relative to the kinetic theory
solution. The kinetic theory result of course also differs
significantly from the ideal Euler hydrodynamic result.

In Figs. 3 and 4 the kinetic theory solutions are compared
to one of the newly developed parton cascade models, ZPC \cite{zpc}.
In that model, only elastic scattering is currently implemented
with a  differential cross section  of the form
\beq
d\sigma/dt=\frac{9\pi\alpha_s^2}{2}(\frac{\mu^2}{s}+1)/(t-\mu^2)^2
\eeq{dsig} 
The total cross section is thus independent of energy,
$\sigma=9\pi \alpha_s^2/2\mu^2$.
However, the more relevant transport cross section needed for the input to
kinetic theory (\ref{etfor}) is
\beq
\sigma_t=\frac{9}{2}\frac{4\pi\alpha_s^2}{s^2}(1+\mu^2/s)\int_{\mu^2}^{s+\mu^2}
\frac{dy}{y}(y-\mu^2)(s+\mu^2-y)
\eeq{sigt}
With this $s$ dependence, 
the average transport cross section changes with time.
Numerical results of ZPC for the time evolution of this quantity
can be parameterized
by 
\beq
\sigma_t(\tau)
= \left\{\begin{array}{ll} 1.74-0.205/\surd\tau\; {\rm mb} & {\rm fig.3}\\
1.80-0.255/\surd \tau \;{\rm mb} & {\rm fig.4}
\end{array}\right.
\eeq{sigt1}

For the physically reasonable values of $\alpha_s=0.47$ and $\mu=3$ fm$^{-1}$,
the transport cross section is so small that even though the parton density is
very high, the scaling Navier-Stokes parameter $\alpha=0.7$ is
small and thus neither Euler nor Navier-Stokes approximations
apply to this problem. The essential point is, however, that
the  ZPC code successfully reproduces the analytic results.

We note that in these examples the ZPC code was
used in the transverse periodic
boundary condition mode to eliminate  effects
of transverse expansion and therefore be directly comparable to the simple
analytic results that were obtained assuming
transverse translation invariance. Numerical results with finite
transverse radius nuclei, show a reduced transverse energy loss
that can also be qualitatively understood as due to a
more rapid decoupling or freeze-out that occurs in that case\cite{rhic97}.
With periodic boundary conditions  the system continues to interact
for a longer time. 
For the conditions shown each event takes about one hour to run on a Sparc 20
from $\tau_0=0.2$ to $\tau=6$ fm.
The numerical results (solid dots) correspond to averaging 20 events.

In Fig 4 the results obtained by 
increasing the initial parton density by two (by decreasing the periodic
transverse area a factor of two)
are shown. These initial conditions  involve one half the transport
mean free path relative to fig.3 and correspond to the scaling Navier-Stokes
parameter $\alpha\approx 1.4$. For the kinetic theory curves the
actual time dependent rates of ZPC using (\ref{sigt1}) were used as input in 
solving (\ref{etfor}).
The accurate reproduction by ZPC of the kinetic theory results
in this case confirms again the numerical accuracy of the ZPC model
in cases far from both free streaming and  hydrodynamic limits.

Finally in Fig. 5, we show results where the initial density was again
doubled to consider a case with $\alpha\approx 3$ that
should closer to the Navier-Stokes
regime. In this case we again find that the ZPC and kinetic solutions
again coincide remarkably well, and in this case  the viscous hydrodynamic
solution approximates much better the kinetic theory evolution.
Physically the price paid for this agreement is the necessity
to increase $\alpha\propto \sigma_t \rho(\tau_0)$ 
by a factor of four relative to the HIJING
estimates. This can be achieved physically in many ways. One way 
is to increase 
the initial mini-jet density by a factor of two (by decreasing the mini-jet
scale $p_0$ in HIJING from 2 to 1 GeV)  and at the 
same time reduce the screening scale
from $\mu\approx 0.6$  GeV to approximately $0.3$ GeV. 
This however is strongly contrary to
pQCD expectation where $\mu\approx g T$ should be  an 
increasing function of the density.In addition such a small
$p_0$ is inconsistent with the data on $p\bar{p}$ at FNAL 
energies\cite{Wangmg}.
One could expect that  inelastic $gg\rightarrow ggg...$ processes,
not yet included into ZPC could increase gluon density somewhat.
However, for our chosen initial conditions, 
$dE_\perp/dy(\tau_0\sim 0.2\;{\rm fm})\approx 0.5 $ TeV in $Au+Au$ 
with $\rho(\tau_0)\approx
20$/fm$^3$ in Fig. 3, the initial energy density
is   $\epsilon(\tau_0)\sim 17$ GeV/fm$^3$, 
instant chemical and thermal equilibrium would imply that
the density would be $\rho_{th}(\tau_0)\approx 2 T_0^3\approx 17$/fm$^3$.
Thus,  the initial conditions are in fact
close to chemical equilibrium in this case and further gluon
multiplication is not likely. It thus appears difficult to approach the
Navier-Stokes regime at RHIC energies. At LHC energies initial densities up to
a factor
of ten higher parton densities may arise 
and thus collective hydrodynamic behavior
should  be more easily be achieved.

Aside from providing physical insight into the possible collective
behavior parton evolution in nuclear collisions,
the  above examples serve well to illustrate how our proposed
analytic test can be applied to newly developed cascade models
as a check of the numerical implementation of kinetic theory.

\section{Summary}

The evolution of the transverse energy per unit rapidity was
computed in kinetic theory for initial conditions of partons
that may arise at RHIC energies. This observable is one of the basic
probes for collective phenomena and is particularly interesting because
Euler hydrodynamic equations predict a factor of two loss
associated with work down as the plasma expands. However, with
pQCD estimates of the density of mini-jets and the screened pQCD cross sections
it appears that local thermal equilibrium may be hard to maintain
even if the initial conditions are assumed to be in boost invariant
local thermal equilibrium. Kinetic theory provides
a microscopic transport theory to evolution far from local equilibrium.
The newly developed parton cascade codes therefore appear to be
essential to compute realistic signatures in nuclear collisions.

In this paper we proposed a simple analytic test of parton cascade models
for idealized  longitudinally boost and
transverse translation invariant initial conditions, and showed 
that this test could be implemented at least in the ZPC\cite{zpc} code.
In the course of applying this test, initial discrepancies between
the analytic results based on eqs.(\ref{etfor}) and numerical ZPC results
were found to be useful in debugging the code.
A similar  test\cite{geiger_test} of  the preliminary version of the
VNI code\cite{vni} 
uncovered an unphysical scattering prescription that 
led apparently to 
anti-work. In the final version of that code that prescription
was  corrected by modifying the 
 low transverse momenta parts of the scattering subroutines so that
similar results to the ZPC test were found.
The fact that analytic results far from the Navier-Stokes domain
could be reproduced by several parton cascade codes
is an important step
in  demonstrating the soundness of the general
parton cascade numerical technology\cite{gcp} being developed
for applications to nuclear collisions at RHIC and LHC energies.
Our main proposal is to subject all codes at least in
the  OSCAR\cite{oscar} repository to such tests.\\[2ex]

Acknowledgments:\\
We are grateful to W. Haxton for support during the INT96/3 Workshop, 
where this work was
initiated and to J. Randrup for support during
during a  summer 97 LBL/LDRD program, where this work was completed. 
Extensive discussions with S. Bass, S. Gavin, K. Geiger, S. Pratt,
D. Rischke, X.N. Wang, 
K. Werner are also gratefully acknowledged.

{}

\newpage
\begin{figure}[h]
\vspace{2.5cm}
\hspace{1.01cm}
\psfig{figure=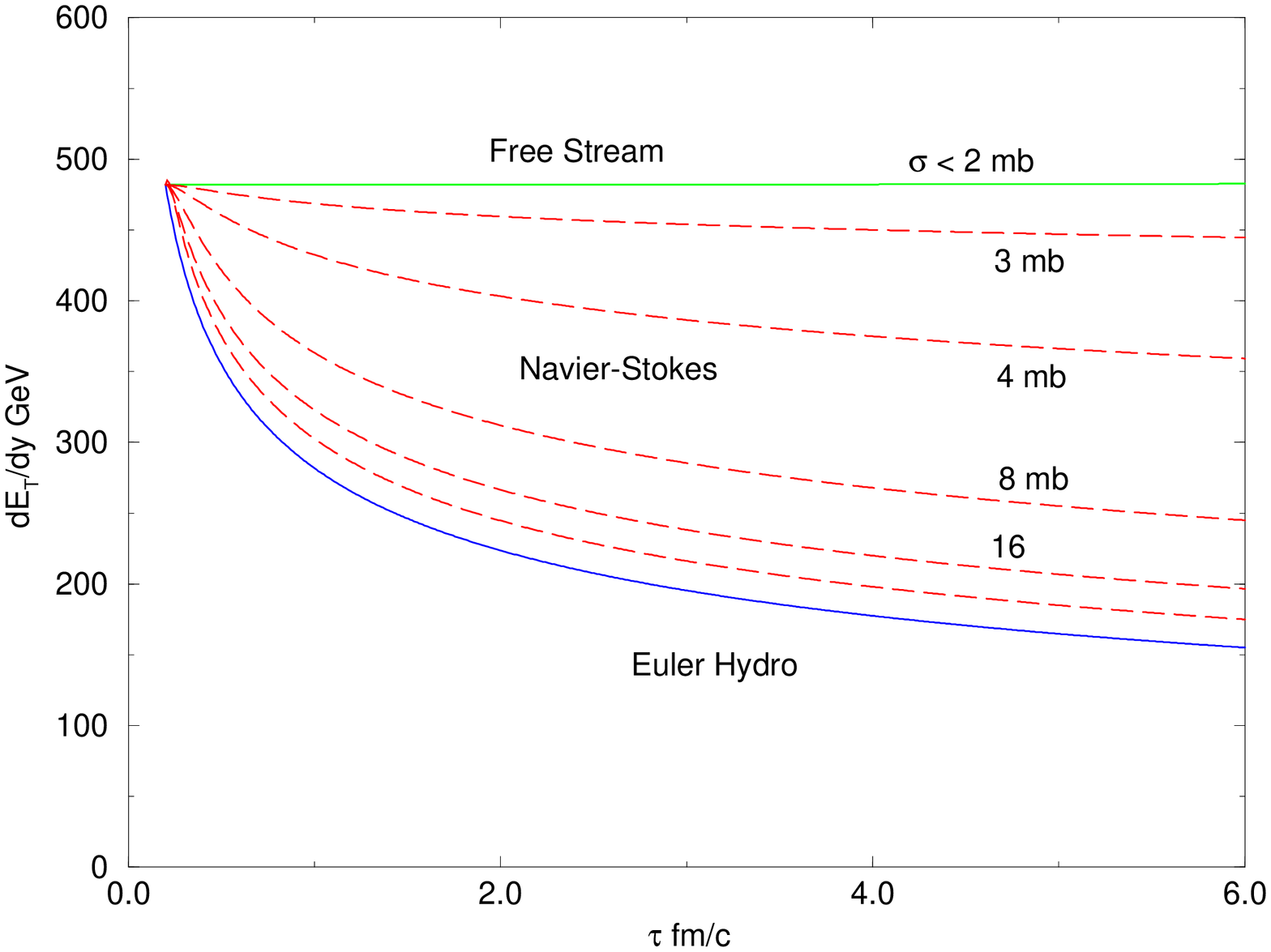,height=3.5in,width=3.0in,angle=-90} 
\vspace{-0.5cm}
\caption{
Comparison of the evolution of the transverse energy per unit rapidity in ideal
Euler hydrodynamics versus Navier-Stokes dissipative hydrodynamics.
The initial conditions were fixed at $\tau=0.2$ fm
to be  $dE_\perp/dy=482$ GeV, $\rho=dN/dy/(\tau A_\perp)=20/fm^3$.
The dependence on the transport cross
section via the viscosity coefficient is shown.}
\end{figure}

\begin{figure}[h]
\vspace{2.5cm}
\hspace{1.01cm}
\psfig{figure=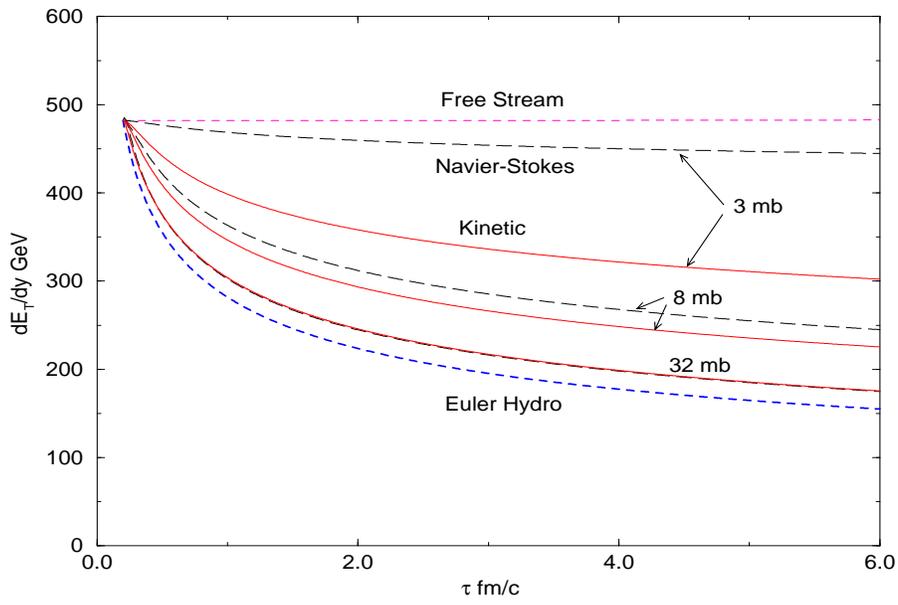,height=3.5in,width=3.0in,angle=-90} 
\vspace{-0.5cm}
\caption{
Comparison of kinetic theory evolution to Navier Stokes and Euler hydrodynamics
for several cross sections given for the same 
initial conditions as in fig.1. Note the slow convergence of
Navier-Stokes to the kinetic theory result as the transport cross section
increases.}
\end{figure}

\begin{figure}[h]
\vspace{2.5cm}
\hspace{1.01cm}
\psfig{figure=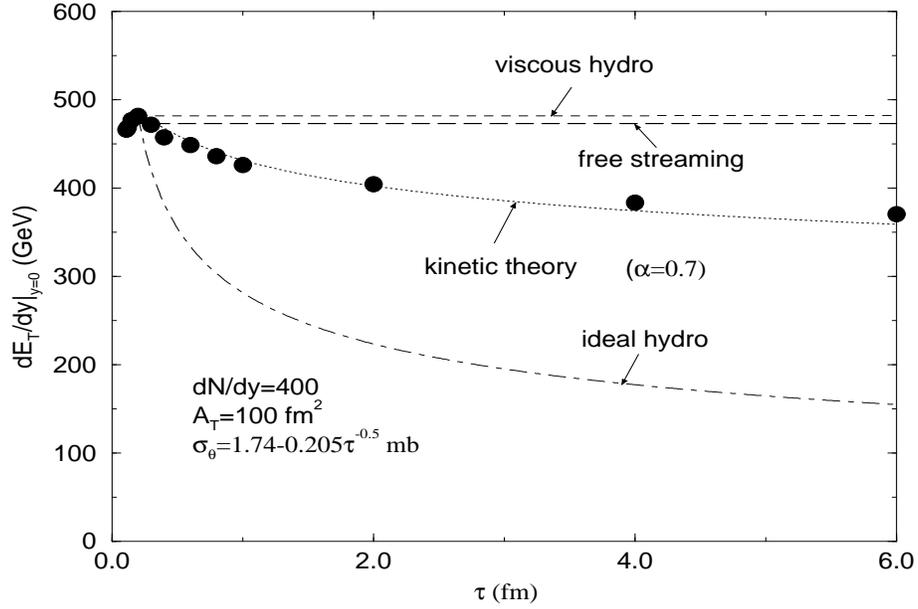,height=3.5in,width=3.0in,angle=-90} 
\vspace{-0.5cm}
\caption{ Comparison of analytic kinetic theory results
to numerical ZPC code\protect{\cite{zpc}} results obtained by averaging
  20 events. A periodic transverse grid of dimensions
$10 $ fm was used. Initially (at $\tau=0.1\;fm$),
  $T_0=500\;MeV$, in an interval $-5<\eta<5$, with
  $\frac{dN}{d\eta}=400$. The screening mass was assumed to be
 $\mu = 3\;fm^{-1}$ with a strong interaction
  coupling constant $\alpha_S=0.47$. The interaction length
for the parton cascade was $0.3\;fm$, and the initial
  mean free path was $\approx 0.3\;fm$.
 The comparison
  starts at $\tau=0.2\;fm$. The good agreement found with the analytic results
confirms the validity of the ZPC algorithm in an interesting
case where the exact results deviate strongly
from free-streaming and dissipative hydrodynamics.
}
\end{figure}

\newpage
\begin{figure}[h]
\vspace{2.5cm}
\hspace{1.01cm}
\psfig{figure=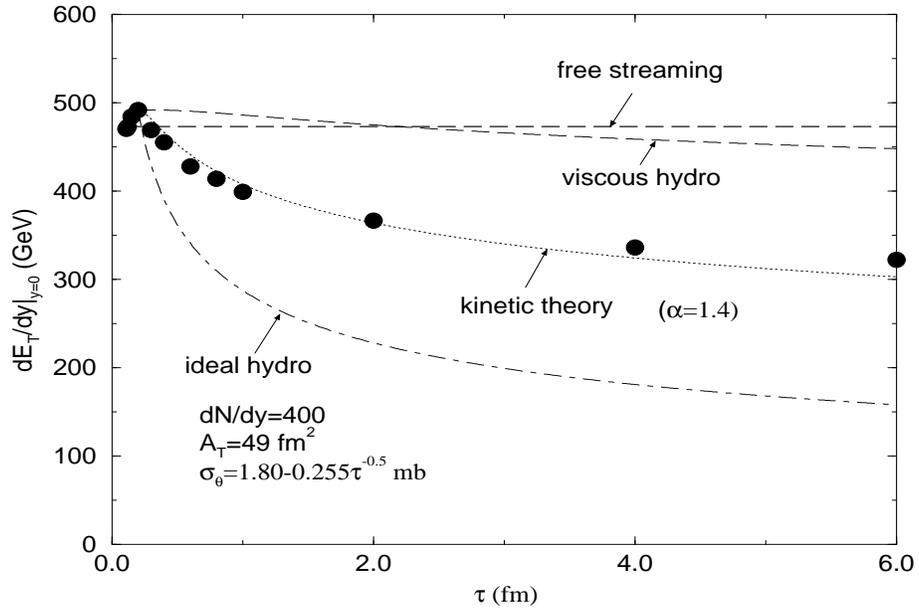,height=3.5in,width=3.0in,angle=-90} 
\vspace{-0.5cm}
\caption{
  Comparison of ZPC results with analytic kinetic theory
for initial conditions with twice as large initial parton density
as in Fig. 3. Here the scaling Navier-Stokes with parameter $\alpha=1.4$,
still deviations strongly from the exact kinetic theory results.
Nevertheless, good agreement with the numerical ZPC model is again found.
}
\end{figure}

\newpage
\begin{figure}[h]
\vspace{2.5cm}
\hspace{1.01cm}
\psfig{figure=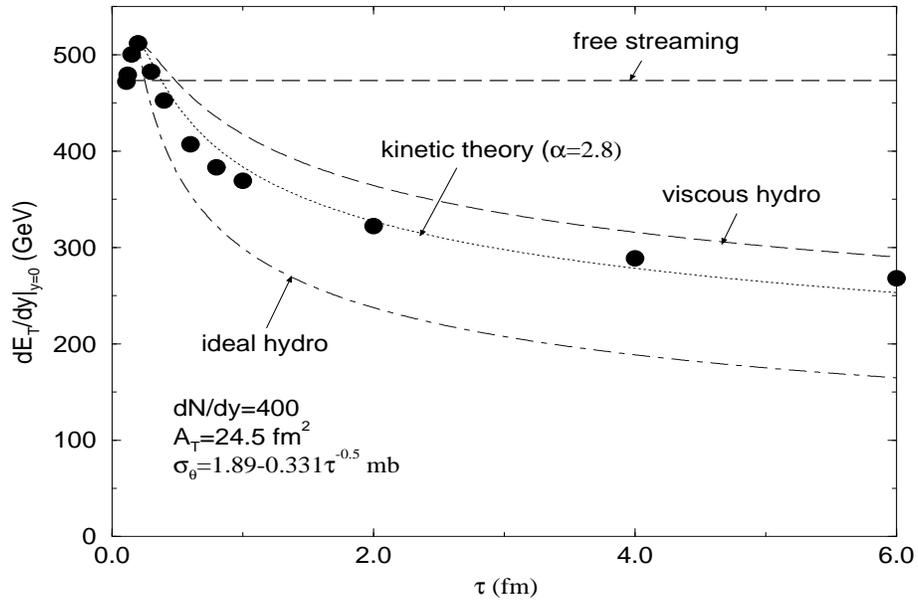,height=3.5in,width=3.0in,angle=-90} 
\vspace{-0.5cm}
\caption{
  Comparison of ZPC results with analytic kinetic theory and
 scaling Navier-Stokes for initial conditions with 
the parameter $\alpha=2.8$. This demonstrates the ability of the ZPC cascade
model to approach the Navier-Stokes dissipative hydrodynamic domain
under extreme initial conditions corresponding to four times the default
HIJING parton density.
}
\end{figure}

\end{document}